\documentclass[times,twocolumn,final]{elsarticle}
\usepackage{float}
\usepackage{breakurl}
\usepackage[hyphens]{url}

\usepackage{framed,multirow}
\usepackage{amsmath}
\usepackage{amssymb}
\usepackage{booktabs}
\usepackage{latexsym}
\usepackage{tabularx}
\usepackage{url}
\usepackage{pifont}
\usepackage[usenames,dvipsnames,table]{xcolor}

\usepackage{hyperref}

\definecolor{newcolor}{rgb}{.8,.349,.1}

\journal{Medical Image Analysis}

\begin{document}


\begin{frontmatter}

\title{
Bias Assessment and Data Drift Detection in Medical Image Analysis: A Survey}

\author[1]{Mischa Dombrowski\corref{cor1}}
\cortext[cor1]{Corresponding author:}\ead{mischa.dombrowski@fau.de}
\author[1]{Andrea Prenner}

\author[1,2]{Bernhard Kainz}

\address[1]{Friedrich-Alexander Universität Erlangen-Nürnberg, Erlangen, 91052, Germany}
\address[2]{Imperial College London, London SW7 2RH, United Kingdom}

\begin{abstract}
Machine Learning (ML) models have gained popularity in medical imaging analysis given their expert level performance in many medical domains. To enhance the trustworthiness, acceptance, and regulatory compliance of medical imaging models and to facilitate their integration into clinical settings, we review and categorise methods for ensuring ML reliability, both during development and throughout the model's lifespan. Specifically, we provide an overview of methods assessing models' inner-workings regarding bias encoding and detection of data drift for disease classification models. Additionally, to evaluate the severity in case of a significant drift, we provide an overview of the methods developed for classifier accuracy estimation in case of no access to ground truth labels. This should enable practitioners to implement methods ensuring reliable ML deployment and consistent prediction performance over time.

\end{abstract}


\end{frontmatter}


\section{Introduction}
\label{sec1}
Advances in deep learning have empowered machine learning (ML) models used in medical imaging to achieve accuracy levels comparable to those of experts across multiple diagnostic and prognostic tasks \citep{banerjee2023shortcuts}. 
ML algorithms have emerged as useful aid for facilitating critical clinical decision-making \citep{jones2023no}. Despite their promising performance, deployments of deep learning models into clinical settings are still limited \citep{salahuddin2022transparency}. One of the main challenges for the adoption of ML solutions in medical imaging, is the development of robust and trustworthy algorithms that are transparent, fair and reliable \citep{saw2022current,salahuddin2022transparency}. 

A growing number of reports highlight concerns regarding the potential of ML algorithms to exacerbate health disparities by perpetuating biases present in the training data \citep{obermeyer2019dissecting,char2018implementing,gianfrancesco2018potential}. This could result in biased outputs across subgroups \citep{banerjee2023shortcuts}, worse performance in underrepresented populations \citep{seyyed2020chexclusion,seyyed2021underdiagnosis} and amplification of health disparities \cite{glocker2023algorithmic}. To counteract these fairness gaps across subgroups, many methods have been developed in recent years to mitigate algorithmic bias in image analysis \cite{wang2020towards,alvi2018turning,madras2018learning,ramaswamy2021fair,li2019repair}. In addition, various papers have focused on increasing transparency regarding the internal bias encoding of medical imaging models \citep{glocker2023algorithmic, glocker2023risk}.


A crucial factor for reliable ML deployment is tracking data drift \citep{kore2024empirical}, which occurs when there's a discrepancy between the data used for model training and the data encountered in clinical practice \citep{sahiner2023data}. Data drift can result from immediate covariate shifts between training and real-world data or evolve gradually due to population changes, such as shifts in patient demographics or the emergence of new diseases.
\cite{duckworth2021using} emphasises that this can lead to unforeseen behaviours, and thereby pose a risk to the safety of patients. It is essential that deployed algorithms are monitored for drift in the populations they were trained for. Monitoring and detecting data drift allows healthcare providers to proactively intervene before the patients' safety is deteriorated. It allows them to assess whether the model should be retrained, taken offline, replaced or re-evaluated \citep{kore2024empirical}. Even though there have been several methods in the ML literature proposed to detect and mitigate the effects of data drift, these methods have not seen broad application in the medical context in general, and especially in imaging systems \citep{sahiner2023data}. Implementing methods to automatically differentiate between malignant and benign shifts is crucial for taking appropriate actions regarding already deployed ML models \cite{rabanser2019failing}.

\begin{figure*}[t]
  \centering
    \includegraphics[width=1.0\textwidth, trim={0 4.5cm 0 4.5cm}]{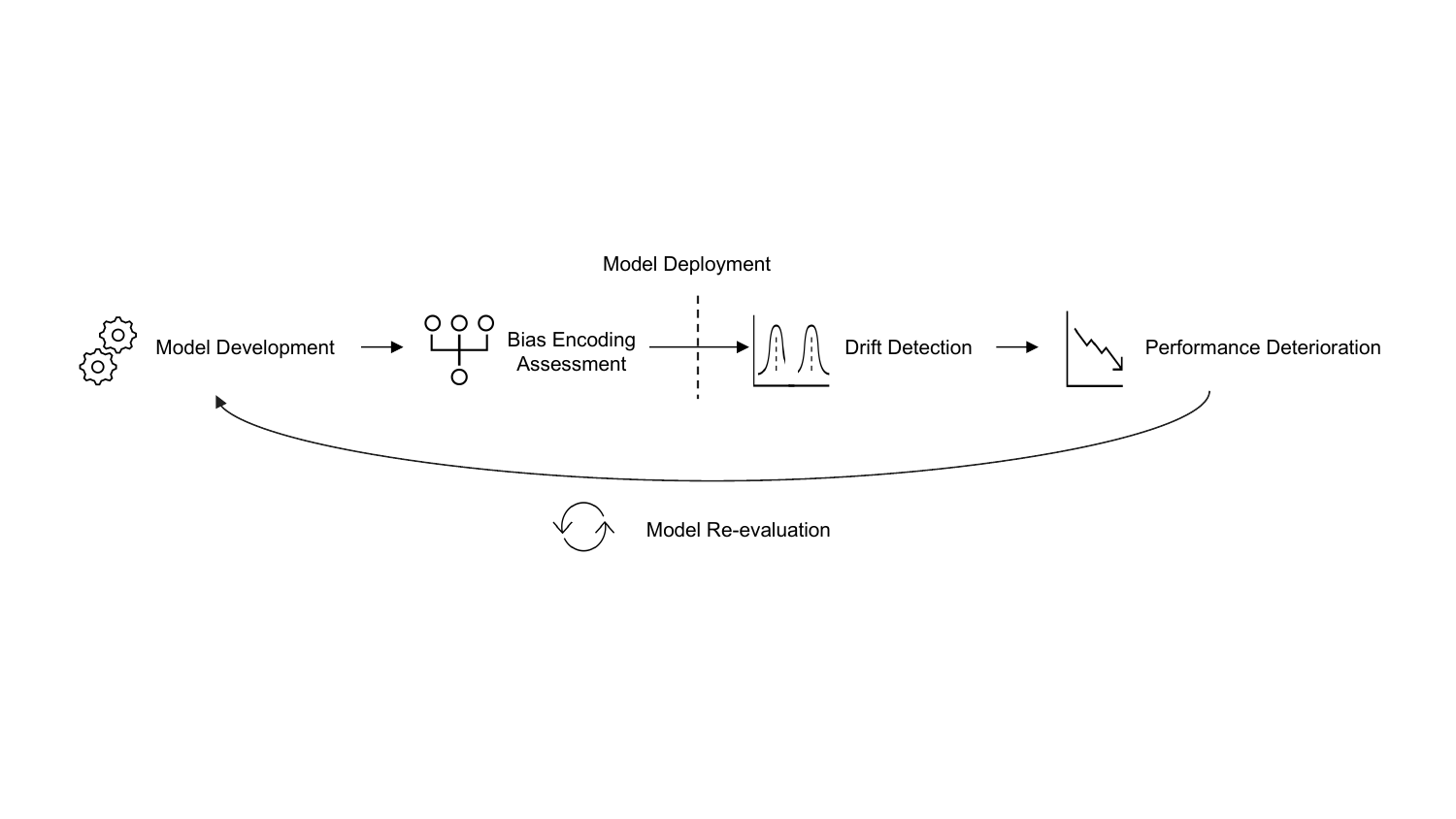}
    \caption{ML Model Reliability Assessment Lifecycle.}
    \label{fig:reliability_assessment_lifecycle}
\end{figure*}

Therefore, the motivation for this survey paper is to offer a comprehensive overview of the available methods  for medical imaging in the field of bias assessment (pre-model deployment) and data drift monitoring (post-model deployment).  Figure~\ref{fig:reliability_assessment_lifecycle} illustrates the life-cycle stages of a ML model that are relevant in terms of reliability assessment and which also represent the focus areas of this paper. 
Our findings are intended to support the translation of fair medical imaging models into clinical practice, ensuring their reliable deployment and sustained performance over time. This survey is specifically aimed not only at machine learning specialists but also at researchers and practitioners from medicine and other fields. As such, we make a conscious effort to avoid domain-specific jargon and describe concepts in simple, accessible terms. 

We focus on several key aspects. First, we clarify the different types of bias that can impact model performance. We then provide a clear distinction between the various types of data drift, an issue that arises when there is a mismatch between the data used to train a model and the data encountered in real-world clinical settings. We also summarise the methods that have been developed to assess bias, ensuring that models are fair and equitable in their decision-making.

Furthermore, we review techniques designed to detect data drift and estimate potential performance degradation, particularly in situations where drift is identified but ground truth labels are not available in a timely manner. In doing so, we also highlight the challenges associated with applying these methods effectively in clinical environments. Finally, we discuss possible future research directions in these areas, pointing to opportunities for further improving the robustness and fairness of medical imaging models.

\section{Key Concepts of Bias}

Bias refers to an estimate of a statistic being systematically different from its population value. If estimates were unbiased on the population level, models would generalise well to other datasets \citep{wachinger2019quantifying}. Similar to \cite{seyyed2020chexclusion, brown2023detecting} we specify biases as ML models exhibiting performance disparities across defined subgroups. Subgroups are  subsets of the population distinguished by specific  attributes, such as, \emph{e.g.}, race, age, and gender \citep{cheong2023causal}. 

\cite{slowik2021algorithmic} outlines two sides to the bias debate: one that focuses on data and another one that focuses on algorithms. We adhere to this distinction and categorize bias into dataset bias and algorithmic bias. 

Even though identifying the source of bias is essential to selecting the appropriate bias mitigation strategies \citep{cheong2023causal}, it still remains a challenge to clearly identify the source of bias given the complexities of high dimensional statistical functions like ML models, the different sources of bias that can contribute to the generation of bias \citep{koccak2024bias} and the innumerable unobservable confounding factors within the data \cite{cheong2023causal}.

\subsection{Dataset Bias} \label{sec:dataset_bias}


A dataset is considered unbiased if its joint distribution $P_{model}(X,Y)$ aligns with the real-world distribution $P_{reality}(X,Y)$. 

Bias concerning the dataset (and therefore a violation of the equality of the joint distributions) can stem from the samples, \emph{e.g.}, images themselves $P(X)$ or from the annotation process. In case of annotation bias, the annotated labels and consequently $P(Y|X)$ are biased \cite{cheong2023causal}. 

\subsubsection{Selection Bias} 
Dataset bias regarding the input $X$ results from selection bias~\citep{wachinger2019quantifying}. Selection bias arises when the participants included in the study do not accurately represent the overall population~\citep{wachinger2019quantifying}, \emph{e.g.}, when a specific subgroup is over- or underrepresented compared to others \cite{banerjee2023shortcuts}. Selection bias in clinical practice is evident when, \emph{e.g.}, radiologic images are frequently collected from just one or a few locations, resulting in a lack of geographic and racial diversity. Additionally, systemic disparities can lead to variations in image quality, with, \emph{e.g.}, Black and Hispanic patients sometimes receiving lower quality and less advanced imaging for similar symptoms in emergency departments, especially in patient cost-driven settings like the USA~\citep{debenedectis2022health}. 
Selection bias can lead to severe class imbalance causing  ML models to predominantly learn from the majority class. As a result, these models tend to achieve high performance metrics for the majority class, but fail to generalise effectively to any minority classes \cite{banerjee2023shortcuts}. 
Similar challenges occur with demographic factors like gender and socio-economic status, where ML models tend to perform more effectively for the demographic groups that are disproportionately over-represented in the training data~\citep{koccak2024bias}. It has to be noted that there is no universally robust training set. Even if the training distribution is representative of the actual testing conditions, the trained model may still perform poorly on certain subgroups. In case of a classification problem, this is caused by the majority and minority population having different classification boundaries ~\citep{slowik2021algorithmic}. 
 
\subsubsection{Annotation Bias}
Label bias, or annotation bias, arises from significant variability among annotators when classifying or delineating regions of interest in diseased areas on data like medical images~\citep{banerjee2023shortcuts}. Annotation shift, where certain subgroups may be systematically mislabelled more frequently than others, can also lead to annotation bias. In case of annotation shift, the model is unlikely to perform well across different subgroups, as the relationship between disease labels and imaging features becomes inconsistent \cite{bernhardt2022potential}.

\subsection{Algorithmic Bias}


Apart from training data, model design plays a crucial role w.r.t. bias amplification. Algorithms are not impartial, and certain design choices lead to fairer prediction outcomes. \citep{hooker2021moving} 
In general, ML systems that focus on minimizing average error have been found to perform inconsistently across significant subsets of the data. Optimizing for the loss averaged over the entire population can easily result in models that perform poorly on specific subpopulations \citep{slowik2021algorithmic}. Model design choices aimed at maximizing test-set accuracy often fail to preserve other important properties, such as robustness and fairness. One key reason these choices amplify algorithmic bias is that fairness often coincides with how the model treats underrepresented protected features. The algorithmic bias a model acquires can be linked to the disproportionate over- or underrepresentation of a protected attribute within a specific category. 
Identifying which model design choices disproportionately increase error rates for protected, underrepresented features is a critical first step in reducing algorithmic harm \citep{hooker2021moving}. \cite{hooker2020characterising, hooker2019compressed} found that compression techniques like quantization and pruning disproportionately impact low-frequency attributes such as age and gender in order to maintain performance on the most frequent features. \cite{jiang2020characterizing} demonstrated that difficult and underrepresented features are learned later in the training process and that the learning rate influences what the model learns. Therefore, early stopping and similar hyperparameter choices disproportionately affect a subset of the data distribution. Some model design choices are better than others regarding fairness considerations. \emph{E.g.} with the widespread use of compression and differential privacy techniques in sensitive areas like healthcare diagnostics, understanding the error distribution is crucial for assessing potential harm. In such settings, pruning or gradient clipping may be unacceptable due to their impact on human well-being \citep{hooker2021moving}. A special case of algorithmic bias is stemming from shortcut learning, which is described in more detail in the subsequent section \ref{sec:shortcut_learning}.



\subsubsection{Shortcut Learning} \label{sec:shortcut_learning}

Shortcut learning or confounding bias emerges when ML models rely on confounding variables to derive predictions \cite{boland2024there}.

Spurious features can lead to shortcut learning, where ML models depend on superficial or irrelevant features. These features are simple for the model to learn but do not generalise beyond the training data where the connection between the label and the spurious feature no longer holds, leading to a drop in performance after deployment. In these cases models may depend on spurious features even when they are less predictive than clinically relevant features \cite{boland2024there}. For instance, ML models relying on portable ICU radiographic markers as proxies for the condition rather than identifying the true underlying pathology for pneumonia prediction \cite{zech2018variable} or pneumothorax detection models relying on inserted chest tubes for prediction \cite{rueckel2020impact}.

Conversely, ML models might use sensitive attributes, such as age, sex, or race, to enhance performance, which could be justifiable when these attributes are correlated with disease risk in the target population. For instance, melanoma is more common in lighter skin tones, breast cancer is more prevalent in women and androgenetic alopecia is more common for men. In these situations, disregarding or removing attribute information could reduce clinical performance~\cite{brown2023detecting}.

\begin{figure}[t]
  \centering
    \includegraphics[width=1.0\columnwidth]{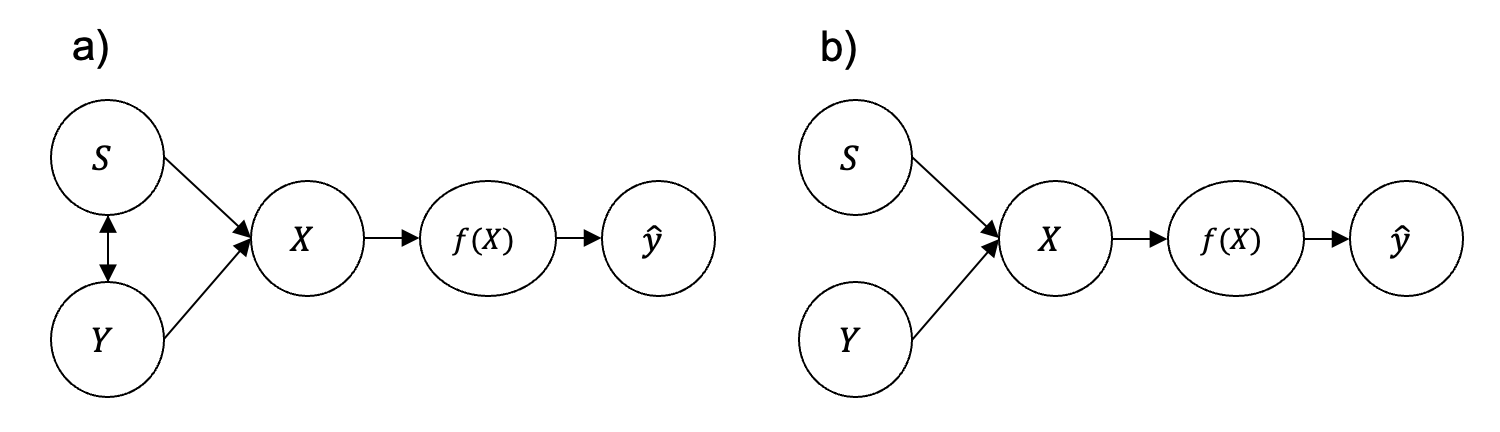}
    \caption{Dependence graphs with a) correlation between sensitive attribute $S$ and label $Y$ and b) without correlation between sensitive attribute $S$ and label $Y$ \citep{dehdashtian2024fairness}.}
    \label{fig:shortcuts_figure}
\end{figure}

Figure \ref{fig:shortcuts_figure}   represents the two distinct scenarios of biases relating to sensitive attributes. In Figure~\ref{fig:shortcuts_figure}a, $Y$ and $S$ are inherently dependent, \emph{e.g.}, an attribute like age can influence both the risk of developing a condition and the appearance of the image. A model might learn to predict the presence of a condition based on the attribute. However, this scenario will introduce a trade-off between performance and fairness. In Figure~\ref{fig:shortcuts_figure}b, $Y$ and $S$ are independent. According to \cite{dehdashtian2024fairness} any observed correlation can be considered as spurious correlation, where these correlations are not considerably beneficial for the performance of the model. In scenario b, the performance of a bias-free model regarding $Y$ to be independent of $S$ is expected.

From a ML perspective, biases can be understood as arising from dependencies between data attributes (confounding variables). The data $X$ depends on the target attribute $Y$ and the sensitive attribute $S$. The aim of bias mitigation is to guarantee that the prediction $\hat{Y}$ is statistically independent of $S$. To grasp the effects of biases, it is useful to distinguish between demographic and non-demographic biases. Demographic biases arise when models perform differently across various demographic groups, which can be defined by attributes like gender, race, or age. Ideally, a bias-free model should exhibit consistent performance regardless of these attributes \cite{dehdashtian2024fairness}. In case of a correlation between $S$ and $Y$, a performance-fairness trade-off is introduced. 

Non-demographic biases, \emph{e.g.}, measurement artifacts (motion artifacts in MRI), are unrelated to demographic factors \cite{spisak2022statistical}. These biases involve spurious correlations that computer vision systems can learn to solve a task. Although these biases are not tied to specific attributes, such attributes may often be identified in various tasks~\citep{dehdashtian2024fairness}.





\subsection{Metrics For Bias Assessment}

To assess the fairness of a trained classification neural network model, the two most common parity metrics are equality of opportunity difference (EOD) and statistical parity difference (SPD) \citep{marcinkevics2022debiasing}. 

In Eq.~\ref{eq:EOD} EOD quantifies the discrepancy between the True Positive Rates (TPRs) of a classifier across the groups of the sensitive attribute $S$ \cite{marcinkevics2022debiasing}:

\begin{align}
EOD = &P_{x,y,S} (\hat{y} = 1 \mid y = 1, S = 0) - \notag \\ &P_{x,y,S} (\hat{y} = 1 \mid y = 1, S = 1),
\label{eq:EOD}
\end{align}

where $y$ and $\hat{y}$ represent the true label and the predicted label, respectively. $S$ represents the sensitive attribute. 

A significant TPR disparity indicates that individuals with a disease within a protected subgroup are not receiving correct diagnoses, at the same rate as the general population, even if the algorithm has a high overall accuracy \cite{boland2024there}. 

SPD is defined in Eq.~\ref{eq:SPD} as the difference between the probabilities of positive outcomes, \emph{i.e.}, predictions made by the model across the groups of the protected attribute $S$ \cite{marcinkevics2022debiasing}:
\begin{equation}
SPD = P_{x,S} (\hat{y} = 1 \mid S = 0) - P_{x,S} (\hat{y} = 1 \mid S = 1)
\label{eq:SPD}
\end{equation}

\subsection{Challenges in Bias Assessment}

The accurate assessment and addressing of bias-related challenges is often limited given that real-world datasets inherently contain biases linked to cohort selection, biases in ``ground truth'' (annotation bias), differences in scanners and protocols or other (un-)known confounding factors within the data or the labels. Moreover, many medical imaging datasets lack sufficient socio-demographic information or representation, making it difficult to thoroughly investigate the full spectrum of bias scenarios, particularly in the context of intersectional analyses \cite{stanley2023flexible}.

\section{Key Concepts of Data Drift} \label{sec:datadrift_concepts}


The implicit assumption underlying all supervised ML techniques is that the training dataset distribution ${P_{tr}(x, y)}$ is the same as the distribution of the data processed by the model post-deployment ${P_d(x, y)}$. In this context, we define  $x \in {R}^n$ as the vector representation of a data item's covariates and ${y}$ its corresponding target variable \cite{dreiseitl2022comparison}. 

Dataset shift occurs when there is a discrepancy in the joint distribution of inputs and outputs (target variables) between the training and deployment stage \cite{quinonero2022dataset} or when data changes slowly over time because of systematic errors or random population shifts. Specifically that is the case when ${P_{tr}(x,y)}$ $\neq$ ${P_d(x,y)}$. Generally, dataset shift can be assumed whenever there are disparities between the distributions of the training and unseen data \cite{moreno2012unifying}. 

In spite of an early effort to standardize the terminology regarding dataset shift \cite{moreno2012unifying}, there are still multiple terms used for the same concept across literature. These terms include domain drift, distributional shift and data set shift among others \cite{sahiner2023data}.

Regarding the term data drift, distinct definitions can be found in literature. \cite{kore2024empirical} and \cite{duckworth2021using} define data drift as ``systematic shift in the underlying distribution of input features'', where $P_{t}$(x) $\neq$ $P_{t+t'}$(x) for probability distribution $P$ defined at time $t$. 
In contrast, \cite{webb2016characterizing} summarise the term drift such that any of the elements of a joint distribution ${P(x,y)}$ might be subject to change over time. Consequently, based on the fact that the joint distribution can be factorized as ${P(x,y) = P(x|y)P(y) = P(y|x)P(x)}$, drift occurs if any of ${P(x|y)}$, ${P(y|x)}$, ${P(y)}$ or ${P(x)}$ change over time.  \cite{quinonero2022dataset} specify dataset drift as circumstance in which the whole joint input-output distribution is non-stationary.

Following the definitions of data drift by \cite{webb2016characterizing} and \cite{quinonero2022dataset}, we will define data drift as any occurrence of change in the joint distribution over time ${P_{t}(x,y)}$ $\neq$ ${P_{t+t'}(x,y)}$. 

Another aspect is that data drift is conceptually different from the traditional task of out-of-distribution detection and anomaly detection~\cite{soin2022chexstray}, as in ~\cite{baugh2023zero, baughimage}. Data drift can manifest as gradual shift in any of the elements of the joint distribution, whereas individual outliers or anomalies might appear without a data drift. In case of drift detection the aim is to intervene at model level (\emph{e.g.}, retrain, removal from production). On the contrary, if out-of-distribution input is identified, e.g. by measuring the difference between the image and its reconstruction ~\cite{muller2022nnood}, the assumption is that the model still performs effectively, but was not accurate for that particular input data~\cite{soin2022chexstray}.

\subsection{Data Drift Implications}

Data drift can lead to malfunction or performance deterioration of ML models \citep{sahiner2023data}. In case of classifiers, the negative effect data drift could potentially have on a classifier's performance is caused by a change in the optimal decision boundary \citep{moreno2012unifying}. Further, it can be the case that the shifted feature distribution may primarily fall into a region where the model performs poorly \citep{duckworth2021using}. 

Therefore, if data drift is detected, a model performance re-evaluation given the current data is necessary \citep{kore2024empirical} and in case of performance deterioration appropriate action needs to be taken, such as retraining the model \citep{duckworth2021using}. 

\subsection{Reasons for Data Drift}

Among the reasons for data drift, two can be considered most relevant: Sample selection bias (Section \ref{sec:dataset_bias}) and non-stationary environments \citep{moreno2012unifying}. 

When there is sample selection bias, the distribution discrepancy between training data and data seen post model deployment arises because the training examples were obtained through a biased method. As a result, they do not reliably represent the operating environment where the classifier will be deployed. The term sample selection bias stems from the fact that training examples are non-uniformly selected from the population during the data collection process thus potentially leading to over-representation of population whose samples are easier to acquire and under-representation or complete exclusion of populations with costly or difficult to acquire samples \cite{moreno2012unifying}. 

Even if the distribution of the training data matches the data post deployment, it may still be subject to drift over time \citep{sahiner2023data}. In case of non-stationary environments, the training environment differs from the deployment environment due to temporal changes in distribution \citep{moreno2012unifying} thereby violating the stationary assumption underlying ML models \cite{cieslak2009framework}.

\subsection{Types of Data Drift}

Following the terminology by \cite{moreno2012unifying} three different kinds of shift can appear. These shifts do not necessarily have to happen independently from each other, but can happen simultaneously. 

When examining dataset shift, the relation between covariates and class labels is highly relevant \cite{moreno2012unifying}. Specifically, a causal and anticausal settings need to be distinguished \cite{castro2020causality}:

\begin{itemize}
    \setlength{\itemsep}{0pt}
    \parskip0pt
    \item Causal: Estimate ${P(y|x)}$ in ${x \rightarrow y}$ problems, where the class label ${y}$ is causally determined by the features ${x}$ (\emph{e.g.},  prediction of labels ${y}$ from medical images ${x}$) \cite{moreno2012unifying,castro2020causality}. 
    \item Anticausal: Estimate ${P(y|x)}$ in ${y \rightarrow x}$ problems, where class label ${y}$ causally determines features ${x}$ \cite{moreno2012unifying,castro2020causality}. 
\end{itemize}

\subsubsection{Covariate Shift}

Covariate shift appears only in ${x \rightarrow y}$ problems, when ${P_{tr}(y|x) = P_d(y|x)}$, but ${P_{tr}(x) \neq P_d(x)}$ \citep{moreno2012unifying}. Thus only the covariate distribution is subject to change between the training and the distribution post deployment, and the model ${P(y|x)}$ remains unaffected. However, ${P_{tr}(x) \neq P_d(x)}$ should not be mistaken as implying that the rule for the prediction of $y$ from $x$ does not need to be adapted to the new covariate distribution ${P_d(x)}$. This is reasoned by the fact that predictions based on finite data may favor simple functions that perform well in regions where ${P_{tr}(x)}$ is high, but not where ${P_d(x)}$ is high \citep{scholkopf2012causal}. 

In case of acquisition shift or domain shifts, the measurement system or method of description is subject to change \citep{quinonero2022dataset}. In medical imaging, domain shift is responsible for potentially harmful disparities between development and deployment conditions of medical image analysis techniques. In this context, acquisition shift relates to variations in the likelihood ${P_{D}(x|z)}$ of an image ${x}$ obtained from a particular domain ${D}$, \emph{e.g.}, its appearance, given the unobserved, latent reality ${z}$ of a patient's true anatomy \citep{hognon2024contrastive}. Domain or acquisition shift can be understood as a change in the mapping function ${x = f(z)}$, where the target variable ${y}$ is dependent on the latent, never directly observable variable ${z}$ \cite{quinonero2022dataset}. Since a change in the mapping function has an effect on the covariate ${x}$, these types of shifts can be seen as subcategory of covariate shift. 

In context of medical imaging, a common cause for covariate shift are different image acquisition devices, \emph{e.g.}, multiple manufacturers of scanners or acquisition protocols in clinical use, leading to varying qualities of images. Patient populations subject to change over time also constitutes covariate shift \citep{sahiner2023data}, \emph{e.g.} when a new disease appears during a pandemic. 

\subsubsection{Prior Probability Shift}

Prior probability shift occurs only in ${y \rightarrow x}$ problems, when ${P_{tr}(x|y) = P_d(x|y)}$, but ${P_{tr}(y) \neq P_d(y)}$ \citep{moreno2012unifying}. The terms label shift and prevalence shift can be used interchangeably for prior probability shift \citep{castro2020causality, garg2020unified}. 

In case of label shift in diagnostic problems where diseases cause symptoms, the optimal predictor might be subject to change, \emph{e.g.}, during a pandemic the probability of a patient having a disease given their symptoms can increase \citep{garg2020unified}. 

\subsubsection{Concept Shift}

Concept shift (Eq.~\ref{eq:cdrift}) occurs when there is a change in relationship between ${x}$ and ${y}$. Specifically it can be defined as \citep{moreno2012unifying}:

\begin{align}
P_{tr}(y|x) \neq & P_d(y|x), \text{ but } P_{tr}(x) = P_d(x) \text{ in }  (x \rightarrow y) \text{ settings }  \notag \\
P_{tr}(x|y) \neq & P_d(x|y), \text{ but } P_{tr}(y) = P_d(y) \text{ in }  (y \rightarrow x) \text{ settings } 
\label{eq:cdrift}
\end{align}

An example of concept shift would be the case if after 2020 certain patterns of patchy ground-glass opacity in chest X-rays might not be labeled as bacterial pneumonia anymore, but as COVID-19 pneumonia \citep{sahiner2023data}.

\subsection{Relevance of Data Drift Detection in Clinical Practice}

Monitoring a ML model post deployment is essential in high-stakes scenarios like healthcare. This ensures that any performance deterioration is detected early and reliable medical care can be provided for patients \citep{kore2024empirical}. Monitoring the model's performance post deployment at the output level by comparing the model's output with the ground truth labels is attractive, but is often infeasible due to a lack of ground truth information in ML stratified workflows~\citep{dreiseitl2022comparison}. 
Automatic data drift detection can alert operators and allow them to take appropriate actions in terms of model retraining, replacement etc. \citep{kore2024empirical}. 
In cases where a drift was detected, but the model's performance did not deteriorate, these insights can be useful in understanding the generalizability of the model to new populations \citep{kore2024empirical}. Further it is of high interest to determine the type of changes that occurred between the training and the post-deployment situation \citep{quinonero2022dataset}. 


\section{Methods for Bias Encoding} \label{sec:methods_bias_encoding}


This section reviews various approaches that uncover or analyse biases in medical imaging models. The aim is to identify inherent biases that compromise fairness and generalisation for disparate subgroups to detect biased models \citep{seyyed2020chexclusion, seyyed2021underdiagnosis}. 

\subsection{Bias Encoding Assessment} \label{sec:bias_encoding_assessment}

\subsubsection{Demographic Confounders}

The following methods provide insights into whether bias was introduced into medical imaging models through the use of sensitive attributes $S$. 

\cite{glocker2023algorithmic}, based on the study of \cite{gichoya2022ai}, where the authors sought to determine whether a deep neural network trained for disease detection might have implicitly learned to recognise racial information, test an intuitive approach of bias detection. Specifically, a deep learning network is trained for a primary task, disease detection in this case, and then the network's backbone parameters are 'frozen'. The prediction layer is then replaced with a new one to learn a new set of weights to solve a secondary task (prediction of a sensitive attribute \emph{e.g.}, race). The features are created by processing the input data through the frozen backbone of the primary task. They then assess the accuracy of the new prediction layer using test data. If the accuracy is sufficiently high, they infer that the two tasks are related and potentially share information. \cite{glocker2023algorithmic} then show that even if this method gives information if sensitive attributes are encoded in the model, the method does not provide insights whether these attributes are used in the prediction process or not. To test their assumption, they used different sets of backbone weights in transfer learning: ImageNet weights, random weights, and weights from a disease detection model trained exclusively on scans from one race. They compared the race detection performance of these models with the original method, which used backbone weights from a disease detection model trained on scans from all races. 
Since the models using ImageNet weights and those using weights from the single-race disease detection model achieved equally high performance in race classification as the original method, they concluded that features learned from the primary task are useful for the secondary task (race detection). However, they could not conclude whether there is a relationship at the output level between disease detection and race classification.

To detect whether protected characteristics (sex, race, age) are encoded by the algorithm and are used in the prediction process (\emph{i.e.} shortcut learning, Sect.~\ref{sec:shortcut_learning}) \cite{glocker2023algorithmic} assess whether the protected attributes align with the learned feature space of the primary task of disease detection. Specifically, they apply dimensionality reduction to the penultimate layer of a trained disease detection DenseNet-121 model. Then they use PCA to capture the main modes of variation within the feature representations. Regarding PCA, these new dimensions (modes) represent the directions of greatest variation in the high-dimensional feature space. Given a model trained for disease detection, the strongest separation of samples with and without disease can be found in the first few PCA modes. Next, they create two-dimensional scatter plots for the first four PCA modes and overlay different types of patient information (disease, sex, race, age). They then plot the marginal distributions arising when projecting the 2D data points against the axes of the scatter plot. Two-sample Kolmogorov–Smirnov tests are used to calculate p-values for the null hypothesis that the marginal distributions of a given pair of subgroups are identical in each of the first four PCA modes. These tests are conducted for all relevant pairwise comparisons involving the presence of disease, biological sex, and race to assess whether particular sensitive attributes are used for deriving predictions (\emph{e.g.}, if there is a statistical significant difference in the marginal distributions of PCA mode 1 in terms of disease, then the marginal distributions of the sensitive attributes for this specific mode should not be statistically significant to ensure that no sensitive attribute is used to derive the disease prediction).

\cite{piccarra2023analysing} apply this framework  to detect biases in generated features in age prediction models based on brain imaging data. Brain-predicted age often serves as indicator of deviations in an individual's brain health from the norm. Their statistical analysis of the model's feature inspection indicates that some features, which are valuable for age prediction, also contribute to distinguishing between racial and biological sex subgroups. Therefore, it can be concluded that the analysed models could lead to significant age prediction disparities among racial and biological sex subgroups.

In a follow-up work \cite{glocker2023risk} use their developed framework (\cite{glocker2023algorithmic}) to analyse a recently published chest radiography foundation model \cite{sellergren2022simplified} for the presence of biases. As part of the analysis, the foundation model is applied on a disease detection downstream task. Dimensionality reduction methods (PCA, t-SNE) in combination with two-sample Kolmogorov-Smironov tests are applied on the high-dimensional features of the penultimate layer generated by the chest radiography foundation model trained on a downstream task. Statistically significant differences are observed in 10 out of 12 pairwise comparisons across biological sex and race in the foundation model studied, demonstrating racial and sex-related bias. The framework by \cite{glocker2023algorithmic} allows practitioners, who might have limited insights into foundation model pretraining, to thoroughly assess bias encoding in foundation models applied to downstream tasks \citep{glocker2023risk}. Given that the foundation model weights were not publicly accessible, the foundation model weights were not updated during the downstream task training. Further analysis on how bias encoding might change when updating foundation model's weights during the downstream task training process is essential.

\cite{brown2023detecting} assess the encoding of sensitive attributes, analyse fairness metrics and introduce shortcut testing (ShortT), which gives insights into the correlation between the encoding of the sensitive attribute and fairness metrics to investigate how shortcut learning might affect model fairness and performance. The underlying assumption is that the influence of sensitive attributes on the model consists of both biological, potentially causal effects that could improve model performance (graph a. in Figure \ref{fig:shortcuts_figure}), and shortcut learning that could be harmful (graph b. in Figure \ref{fig:shortcuts_figure}). Even though both scenarios introduce bias, \cite{brown2023detecting} only refer to shortcut learning when a sensitive attribute is used as confounder, which does not significantly enhance performance, but impacts fairness. To assess the encoding of sensitive attributes, age in their analysis, they trained a model for disease prediction, froze all weights in the model backbone and then trained a predictor for age using a mean squared error (MSE) regression loss. The performance of the transfer model is measured using the Mean Absolute Error (MAE), which serves as an indicator of the age-related information captured by the final layer of the feature extractor. Lower MAE values correspond to more accurate age predictions, indicating stronger age encoding. To quantify fairness of the model's disease predictions in relation to age, \cite{brown2023detecting} calculate the separation metric. Separation was defined by fitting two logistic regression models to the binarised model predictions: one for patients with the condition and one for those without it. This approach is equivalent to modelling the effect of age on the TPR and the False Positive Rate (FPR). The separation value was determined by averaging the absolute values of the two logistic regression coefficients. A score around 0 indicates no monotonic relationship between age and model performance, while higher scores imply that TPR and/or FPR systematically vary with age. To assess the degree of age encoding on model fairness, they use a multitask learning model by adding an age prediction head to the base model (disease prediction model) and scale the gradient updates from the age prediction head. 
For each value of gradient scaling, they compute the model disease prediction performance, MAE of the age prediction and separation fairness metric. Shortcut learning is then indicated by a significant correlation between age encoding and separation fairness metrics (computed via Spearman correlation coefficient). The proposed framework represents a feasible approach for analysing the impact of shortcut learning since the analysis only involves the addition of a demographic prediction head to the base model. Practitioners need to carefully consider which fairness metric to select (besides separation, \emph{e.g.}, demographic parity (independence) could be selected), and whether the gradient intervention consistently modifies the encoding of the protected attribute before evaluating its relationship with model fairness \cite{brown2023detecting}.

\subsubsection{Non-Demographic Confounders}


\cite{boland2024there} use Prediction Depth (PD) to detect shortcut learning after adding synthetic shortcuts to the training data. PD measures the example difficulty by the number of layers a model requires to make a final prediction. \cite{boland2024there} link PD to shortcut learning and the simplicity bias in neural networks, which was originally proposed by \cite{murali2023beyond}, who showed that shortcuts are harmful when they are simpler than the relevant features. To test their method they train two binary classification models, one with no shortcuts and the other one with synthesised  shortcuts (small red squares, vertical, semi-opaque curved lines added to images), which are perfectly correlated with one class. Both models are evaluated on the same test set, with the shortcut balanced equally between the two classes. To verify the hypothesis that shortcuts reduce PD, the Welch's t-test is applied to test if there is any statistically significant difference in mean PD between the two models. One dataset in their analysis is the CheXpert medical imaging data set, where they also introduce synthetic shortcuts to the data. Since the method was only tested on synthetically created shortcuts, its effectiveness on natural shortcuts, such as site-specific shortcuts in multi-site data, remains to be evaluated. Additionally, the current method has a limitation in that it always requires a dataset without shortcuts as a reference.


\subsection{Confounding Bias Detection Through Causal Inference}

\cite{wachinger2019quantifying} introduce a method for distinguishing between causal and confounded relationships using causal inference. Confounding w.r.t. the neuroimaging dataset used in their analysis refers to imaging site-specific information being implicitly encoded by the model. Notably, the confounders are treated as unknown, latent variables in their method, which is beneficial in complex neuroimaging studies (and also in other medical imaging datasets) where the assumption of causal sufficiency (\emph{i.e.}, knowing all confounding variables) is often violated in practice. Their method aims to assess whether $X$ is more likely to cause ${Y}$, or if an unobserved random variable ${Z}$ is the underlying cause of both. Their method leverages the algorithmic Markov condition (AMC), which asserts that if $X$ causes $Y$, the factorisation of the joint distribution ${P(X,Y)}$ in the true causal direction will have a lower Kolmogorov complexity than the reverse, anti-causal direction. To approximate Kolmogorov complexity, which is not directly computable, the minimum description length ($L$) is applied. The factorisation of the causal scenario ${P(X,Y)_{ca} = P (Y\mid X) P (X\mid Z) P(Z)}$ is represented by a linear regression model, whereas the factorisation of the confounded scenario ${P(X,Y)_{co} = P (Y\mid Z) P(X\mid Z) P(Z)}$ is estimated by probabilistic PCA. To compare the causal with the confounded model $\Delta(X,Y) = L_{co}(X,Y) - L_{ca}(X,Y)$ is computed. If the causal model provides a better fit to the data than the confounded model then ${\Delta(X,Y) > 0}$. It has to be noted that inferring causality from observational data is difficult because the true causal effects and their magnitudes are unknown, making quantitative evaluation impossible.

\subsection{Representative Datasets For Bias Assessment}

To objectively analyse the impact of biases on medical imaging models without the limitations of real-world datasets (\emph{e.g.}, unknown confounding factors), \cite{stanley2023flexible} propose SimBA, a versatile framework for generating synthetic neuroimaging data with controlled simulation of disease, bias, and subject effects. The synthetic images are generated by applying non-linear diffeomorphic transformations to a template image $I_{T}$ that represents the average brain morphology. The non-linear transformations for disease and bias are spatially localised deformations to $I_{T}$ whereas the subject morphology is generated through a global non-linear transformation. These effects and deformations were derived from PCA-based generative models of these non-linear deformations. Each effect represents a specific degree of morphological variation within the typical range of inter-subject human brain anatomy. This method allows to generate synthetic datasets with controlled bias effects, but it has to be noted that expert knowledge is required to define the brain regions for the localised deformation for the disease and bias effects. Further it needs to be highlighted that the introduced biases are not designed to replicate any particular real-world sociodemographic subpopulation, as the imaging features that introduce bias within these groups are often complex, interrelated and/or unknown. Instead, the simulated biases create hypothetical subgroups within a dataset that exhibit specific confounding features, which may lead to shortcut learning in medical imaging ML models. However, the synthetically biased datasets provide an opportunity to test bias mitigation strategies.

Apart from these synthetically generated biased/unbiased datasets, \cite{glocker2023algorithmic} use strategic resampling with replacement to create balanced test sets that represent the population of interest. This aims to not replicate dataset bias in the test set (due to random splitting of the original dataset) and to allow an unbiased estimation of performance on different race subgroups. Specifically, they use resampling with replacement to create race balanced test sets while controlling for age differences and disease prevalence for each racial group.

\subsection{Simulated Bias Encoding}

\cite{stanley2024towards} extend their research on the SimBA framework \citep{stanley2023flexible} by using this tool to study bias manifestations and the effectiveness of bias mitigation techniques. They generate counterfactual neuroimaging datasets with three bias scenarios (No Bias, Near Bias, Far Bias with Near and Far indicating the proximity to the voxels representing disease). By incorporating unequal proportions of biased brain images for the disease class and non-disease class (70\% vs. 30\% containing bias) they incorporate the possibility of the model to use bias as shortcut for predicting the disease. At the same time they ensure that subject and disease effects are similar for each bias group (preventing unwanted additional source of bias) to enable a controlled evaluation of bias. \cite{stanley2024towards} then use these synthetically generated images to assess the efficiency of various bias mitigation strategies (reweighing, unlearning, group models). The SimBA framework represents an interesting approach to test models for bias encoding, but the bias scenarios that can be generated by this framework are limited to scenarios, where biases are represented by localised spatial deformations, or are related to intensity-based simulated artifacts. Real-world medical imaging data can be more complex and inherently contain numerous confounding and interacting biases.

\section{Methods for Data Drift Detection} \label{sec:methods_datadrift}

This section provides an overview of methods developed for data drift detection in the imaging domain in Sect.~\ref{sec:methods_datadrift_imaging} and on methods which have already been successfully applied to medical imaging workflows in Sect.~\ref{sec:methods_datadrift_medical_imaging}. 

\subsection{Drift Detection in High-Dimensional Data} \label{sec:methods_datadrift_imaging}

\cite{rabanser2019failing} base their analysis of shift detection between imaging source (training) and target data (artificially shifted data) on dimensionality reduction methods combined with two-sample tests. They used the MNIST and CIFAR-10 imaging datasets for their analysis. 

Table~\ref{tab:rabanser_methods} gives an overview of the dimensionality reduction techniques that they applied on the source and target data and the respective statistical test applied on the dimensionality-reduced latent space to detect shifts in source and target data distributions. The label classifier dimensionality-reduction method is based on the ResNet-18 architecture trained on the source data to predict the classes present in the datasets. Specifically, they use the Softmax outputs and binary  predictions after thresholding for further statistical testing. They trained a domain classifier to distinguish source from target data. 

\begin{table}[H]
\centering

 \scriptsize
   \label{tab:table1}
   \setlength{\tabcolsep}{2pt}
    \begin{tabular}{c|c}
      \textbf{Dimensionality Reduction}&\textbf{Statistical Test}\\
    \hline
    PCA  &   Multivariate Kernel Two-Sample Tests: MMD,  \\
         &   Multiple Univariate KS tests\\
    \hline
    Sparse Random Projection 	& Multivariate Kernel Two-Sample Tests: MMD,  \\
         &   Multiple Univariate KS tests\\
    \hline
    TAE/UAE	& Multivariate Kernel Two-Sample Tests: MMD,  \\
         &   Multiple Univariate KS tests\\
    \hline
    Label Classifier & Multivariate Kernel Two-Sample Tests: MMD,  \\
         &   Multiple Univariate KS tests, \\
         &   Chi-Squared Test (thresholded classifier predictions) \\
    \hline
    Domain Classifier & Binomial testing  (${H_0: accuracy = 0.5}$) \\
    
    \end{tabular}
    \caption{ Overview on methods applied by \cite{rabanser2019failing} on MNIST and CIFAR-10 imaging data with simulated shifts; TAE = trained autoencoder, UAE = untrained autoencoder, Maximum Mean Discrepancy = MMD, Kolmogorov-Smirnov = KS}
    \label{tab:rabanser_methods}
\end{table}


To compare the differences between the source and shifted dataset after reducing their dimensions, the authors used several statistical methods. They applied Multivariate Kernel Two-Sample Test using Maximum Mean Discrepancy (MMD) to compare the overall distributions. Then, they used permutation testing to assess the significance of these differences. They also ran individual tests on each dimension separately using the Kolmogorov-Smirnov (KS) test. Since they performed multiple tests, they adjusted the results using Bonferroni correction to avoid errors from multiple comparisons. Additionally, they used Pearson’s Chi-squared test for the binary label classifier to check if the class labels were distributed similarly between the source and target datasets. Lastly, for the domain classifier trained to discriminate target from source data, to check if the classifier’s performance was by chance, they used binomial testing, with a null hypothesis that the classifier’s accuracy would be 50\% (random guessing).

On their simulated shifts affecting both covariates and label proportions, \cite{rabanser2019failing} find that the Softmax outputs from the trained label classifier combined with multiple univariate KS testing yields the best performance in shift detection, followed by multivariate-testing of the untrained Autoencoder (UAE) latent space. 
It has to be highlighted that the number of target (shifted) data samples has a substantial influence on the accuracy of the shift detection. 
This has to be taken into consideration when applying the shift detection methods in practice. Consequently, selecting an appropriate rolling window time frame is essential to have a sufficient amount of data samples to compare with the reference (training) data. Since the shift detection based on Softmax outputs combined with KS testing can be easily added on top of an already trained classifier, this method constitutes a feasible practical solution \cite{rabanser2019failing}.  

Based on the findings of \cite{ovadia2019can} it can be concluded that entropy scores and prediction confidence are ineffective for identifying data shift in imaging datasets. When tested on out-of-distribution (OOD) data (based on a class excluded from the training set in the MNIST imaging dataset), most calibrated models exhibited low entropy and high confidence in their study, meaning they were confidently incorrect when predicting entirely OOD data. Consequently, changes in these two metrics do not provide a reliable indication of data drift leading to increased uncertainty, as measured by entropy scores and prediction confidence. Concerning the Expected Calibration Error (ECE) they found that ECE increased as the shift intensity grew. However, since ECE depends on ground truth labels, using this measure to detect drift is not practical for automated drift detection in clinical settings.

\subsection{Methods for Drift Detection in Medical Imaging} \label{sec:methods_datadrift_medical_imaging}

Even though domain adaptation, which becomes relevant in case of dataset shift (domain shift), has gained significant attention over the last years \cite{guan2021domain}, the methods on mere detection of data drift in the medical imaging context are still limited \cite{sahiner2023data}. This section will provide an overview of successfully applied drift detection methods for medical imaging workflows. 

\subsubsection{Image Characteristics and Classifiers}
\cite{kore2024empirical} applied dimensionality reduction using a trained Autoencoder alongside Softmax outputs from a model trained on the source data (Blackbox Shift Detection, BBSD). Also, the combination of trained Autoencoder and BBSD was employed to detect data drift in chest X-ray images, drawing parallels to the work of \cite{rabanser2019failing}. 
The trained Autoencoder is a TorchXRay Vision AutoEncoder (TAE)\citep{cohen2022torchxrayvision} and the BBSD model is a TorchXRayVision model. They applied the MMD statistical test on the resulting embeddings from these three dimensionality reduction methods to identify statistical differences between the source and target data. The shift detection methods were tested on temporal imaging data, which initially constituted natural shifts (such as the introduction of COVID-19 in 2020, which caused a prevalence shift), and subsequently were synthetically shifted data (stratification to create random samples based on the overall population and then enriching with specific categories, \emph{e.g.}, male sex, to study specific population shifts). The shift factors studied were institution, sex, patient age, patient class, and ICU admission status. The authors found a correlation between the sensitivity of drift detection and the magnitude of the synthetic drift, with the TAE+BBSD combination proving more sensitive than TAE alone, and BBSD being nearly as sensitive as the combined method. Future research could explore whether these drift detection methods also work on shifted imaging data with regards to patient attributes race and ethnicity. 
Additionally, only a small subset of covariates was stratified for the synthetic shift, and the potential influence of unmeasured confounding variables that could affect the shift's intensity must also be considered.

\cite{koch2024distribution} employed a domain classifier to distinguish between the source and shifted datasets. They also performed MMD tests, where the kernel 
$k$ was parameterized by a neural network, on features extracted by a neural network trained separately on the source and shifted training sets. Additionally, they analysed the softmax prediction outputs of a classifier trained to predict the $C$ classes from the source dataset, using multiple univariate Kolmogorov-Smirnov tests. With this  \cite{koch2024distribution} tackled the challenge of detecting clinically significant distribution shifts in retinal imaging data gathered during diabetic retinopathy screenings from multiple hospitals, each with diverse demographic populations. 
Similar to the approach taken by \cite{kore2024empirical}, they simulated subgroup shifts based on patient characteristics such as sex, ethnicity, image quality, and the presence of comorbidities. For example, a shift in patient sex was simulated by including only images from female patients in the target distribution. Their experiments demonstrated that classifier-based tests consistently and significantly outperformed the other methods. It is important to note that in clinical practice, splitting data into source and target sets to train a classifier for distinguishing between the two may be challenging, as potential shifts are often unknown in advance, making it difficult to anticipate which distributional shifts might occur in the future.
Additionally, crucial covariates that define relevant subgroup characteristics are often unmeasured or unidentified. Therefore by only controlling for a fraction of possible covariates, the shift might not be originating from the shifted feature, but from an altered composition of (unobserved) covariates. 

\subsubsection{Adding Patient Metadata For Drift Detection}

As opposed to the works of \cite{kore2024empirical} and \cite{koch2024distribution}, \cite{merkow2023chexstray} not only used image appearance as relevant feature for drift detection, but also integrated DICOM metadata (\emph{e.g.}, patient demographics, image formation metadata, image storage information) and predicted probabilities into an aggregated metric to detect temporal data drift in X-ray datasets using Rolling Detection Windows. To encode image appearance-based features and derive the latent space representation, a variational autoencoder (VAE) is used. To verify the statistical significance of the drift for continuous features (\emph{e.g.}, age, latent encoding from VAE) Kolmogorov-Smirnov tests are used and for categorical features (\emph{e.g.}, sex) Chi-Squared tests are applied. \cite{merkow2023chexstray} simulated two drift scenarios: (1) performance degradation -- randomly populating the stream of original data  with hard data by only focusing on samples where the classifier exhibits low confidence for individual labels -- and (2) clinical workflow failure -- introducing lateral view images which the model was not trained on; adding paediatric data that typical ML models are not authorised to report on. They could show that statistical significant variation in model inputs and outputs can be useful indicators of potential declines in model performance. 

\subsubsection{Drift Detection Using Layer Activations}

The work of \cite{stacke2020measuring} focused on quantifying the magnitude of domain shift by measuring shift in the learnt representation space of Histopathology imaging data. Under the assumption that in a well-trained model a convolutional layer will focus on image features that are relevant to the specific task and irrelevant features are discarded, \cite{stacke2020measuring} developed a metric (the representation shift metric $R_{l}$, which is defined for each layer $l$) that analyses the filter activations in each layer of a given CNN model. Even if the differences between the two domains appear minor in the image space, variations in training data statistics can lead to significant discrepancies in the internal model representations. First, all samples from the source and target dataset are passed through the model and in each feature map for each layer the representations are averaged over the dimensions $h$ and $w$: $c_{lk}(x) = \frac{1}{h} \frac{1}{w} \sum_{i,j}^{h,w} \varphi_{lk}(x)_{i,j}$, where $\varphi$ represents the layer activation at each layer $l$ and filter $k$ produced by an input sample $x$. Next, the representation shift $R_{l}$ is derived by calculating the discrepancy between the distributions of $c_{lk}$ between the source and target dataset and then taking the average over all filters, which results in one $R_{l}$ per layer. In their study \cite{stacke2020measuring} evaluated three discrepancy metrics: Wasserstein distance, Kullback-Leibler divergence, and Kolmogorov-Smirnov statistic. 
If the data samples from the source and the target domains are statistically similar in the image space, or if the CNN maps both datasets to comparable representations, then the filter responses should be similar resulting in small $R_{l} = (p_S, p_T)$. 
They found a strong negative correlation between the $R_{l}$ metric and model performance, depending on the layer, model, dataset and discrepancy metric used. The proposed metric for measuring domain shift in the learned representation space provides insights for identifying data the model has not encountered during training. Since the metric does not rely on annotated data, it can be used as a straightforward initial test to assess whether new data (\emph{e.g.}, histopathology images from a different scanner) is properly handled by an already trained model, \emph{e.g.}, whether the learned feature representation applies to the new data. Additionally, it can be used to monitor a deployed model in a clinical setting, where changes in data may introduce image variations that the model cannot handle, which would be detected by an increase in representation shift.

\subsubsection{Prevalance Drift Detection}
\cite{roschewitz2023automatic} introduced Unsupervised Prediction Alignment (UPA), which was originally designed to recover the desired sensitivity/specificity trade-off in the case of acquisition shift. UPA can also be applied to detect prevalance shift. This involves using linear piecewise cumulative distribution matching to align the prediction distribution from the unseen dataset with a reference prediction distribution. In the case of prevalance shift detection, the mean absolute difference between the original (shifted) and aligned predictions is calculated -- after applying UPA to match predictions of reference set with fixed prevalance. When there is no prevalance shift present in the data, the difference should be 0. A limitation of this method is that it only detects shifts in prevalence, reducing its clinical practicality as it would need to be used alongside other drift detection techniques.

\subsubsection{Shift Detection Across Datasets}
To detect and quantify shifts in musculoskeletal radiographs (MURA) and chest X-ray datasets \cite{guo2023medshift} propose MedShift. In this approach, the source datasets, Emory Chest X-rays and Emory MURA, are used as the baselines, while CheXpert \citep{irvin2019chexpert} and MIMIC \citep{johnson2019mimic} in case of the Emory Chest X-rays base data and Stanford MURA \citep{rajpurkar2017mura} in case of Emory MURA base data serve as the target (shifted) datasets. The source dataset is divided into its respective disease classes, and for each class, anomaly detectors using Cascade Variational Autoencoders are trained to capture in-distribution variations. Then, the trained anomaly detectors are applied to the classes in the target dataset to acquire anomaly scores. The anomaly scores (${S = L_{rec} + S_{dis}}$) are derived from adding the reconstruction error ($L_{rec}$) and the probability of being the anomaly class ($S_{dis}$, which is obtained from the discriminator). Based on the anomaly scores the data for each class in the target dataset is clustered into groups in an unsupervised way. To test the \emph{shiftiness} of each of the class groups in the target dataset, a multi-label classifier is trained on the source data set and then applied to the target data classes to assess the generalisability on the target dataset without removing anomalies. Then the groups with the highest anomaly scores are dropped one by one and the corresponding class-wise classification performance is recorded. The variation in performance serves as an indicator of the degree of shift within the specific group. The authors showed that the trained model classification accuracy is rising on the external dataset after removal of shifted data items. The proposed method is efficient to detect variability between datasets when they exhibit the same classes and could potentially be used to assess if a given dataset is similar to the training set used for model training. In their analysis \cite{guo2023medshift} found that higher anomaly groups are originating from variations in positioning, noise and image quality. Therefore, further investigation is needed to determine the specific types of shifts to which this method is sensitive.

\section{Evaluating the Impact of Data Drift}
\label{sec:data_drift_harmfulness_assessment}

Distribution shifts can cause a significant decline in the performance of statistical models, but not all shifts are harmful. Therefore, distinguishing benign shifts from harmful shifts is essential \cite{rabanser2019failing}. Since ground truth to evaluate model performance might be difficult to obtain in a timely manner or might not be available in general \cite{kupinski2002estimation}, methods to evaluate model performance without access to ground truth labels are essential. The methods in ~\ref{sec:drift_severity_assessment} and ~\ref{sec:classifier_accuracy_estimation} have been evaluated on image data, but have yet to be tested on medical imaging data.

\subsection{Drift Severity Assessment} \label{sec:drift_severity_assessment}
\cite{rabanser2019failing} propose evaluating the accuracy of samples predicted as belonging to the shifted data by a domain classifier with high confidence. This classifier is trained to distinguish between source and shifted data. The predictions for these shifted samples are then compared to their true labels. If the predictions are inaccurate, it indicates that the shift is harmful. While this approach still requires ground truth labels, it reduces the labelling effort by only needing labels for the samples predicted most confidently to belong to the shifted data domain.

\subsection{Classifier Accuracy Estimation}
\label{sec:classifier_accuracy_estimation}
To estimate accuracy of an unlabelled data set, \cite{deng2021labels} introduce the Automatic model Evaluation (AutoEval) method. They estimate the accuracy by training a Linear Regression model and Neural Network Regression model to predict the accuracy ($\hat{acc} = A(f)$, where $A$ = regression model and $f$ = features of the respective data set). In case of the Linear Regression model, $f$ represents the Fr\'echet distance measuring the domain gap between the features of the penultimate layer of the classifier of the original data set and the test data set ($f_{linear} = $ \(\| \)  $\mu_{ori}- \mu$ \( \|^2_2 \) $+Tr (\sum_{ori} + \sum - 2(\sum_{ori} - \sum) ^{\frac{1}{2}}))$. For the Neural Network Regression model, $f$ represents the Fr\'echet distance and additionally mean vector and covariance matrix of the test set features ($f_{neural} = [f_{linear};\mu;\sigma])$. 
To train the Linear Regression and the Neural Network Regression model (thereby learning the mapping function $A$ from distribution shift to accuracy), synthetic meta sets based on the original data set are constructed through visual transformations resulting in a diverse data distribution but same label space as original data set. They found that Neural Network regression achieves better results than the Linear Regression model (measured in RMSE between predicted accuracy $\hat{acc}$ and true accuracy). The mapping function $A$ can then be applied to $f$ of future natural variations of the original datasets to estimate the accuracy. It has to be highlighted that this method is highly dependent on the meta sets, which are used to train the model. The method is built on the assumption that variations in real-world samples can be approximated through image transformations within the training meta-set. If future data exhibits very special conditions or patterns or has an entirely different set of classes, then this method might not work. 

For accuracy assessment, \cite{press2024entropy} use an  Entropy minimisation (EM) approach by observing how the embeddings of input images are affected as the model is optimised to reduce entropy. Their method is based on their observation that there exists a correlation between the accuracy and the Silhouette score of the clustered embeddings, implying that in case of a high accuracy, the input data is already clustered well. Given that EM works by clustering its inputs, EM is not expected to make significant changes to an already well-clustered set of embeddings - resulting in only a few label flips. On the contrary, for a dataset with low accuracy, the initial image embeddings are likely to be poorly clustered. This results in the EM algorithm making substantial changes to the embeddings, leading to numerous label flips. Following these considerations, they introduce the Weighted Flips (WF) measure, which takes into consideration the label flips ($1_{flip} (i)$) and the classifier's initial confidence ($c_{i}$) in its predictions ($WF = \sum_{i}1_{flip} (i) c_{i}$). Images that are initially classified with high confidence that later flip should have a greater impact than those classified with lower initial confidence. Utilising $k$ pairs of weighted flips and accuracy ($(WF, accuracy)_{k}$), a weighted-flips-to-accuracy function $f$ is derived through interpolation. With this weighted-flips-to-accuracy function $f$, the accuracy of an unlabelled dataset can be estimated based on the WF measure of this data set.

\cite{guillory2021predicting} introduced the method of differences of confidences (DoC) for accuracy prediction on unseen image distributions. Specifically they propose DoC as: $DoC_{B,T} = AC_{B} - AC_{T}$, where $AC_{B}$ represents the average confidence of the source distribution and $AC_{T}$ represents the average confidence of the target distribution with respect to classes present in both target and source distribution. Additionally, they also propose the metric difference of average entropy (DoE) of a model's output probabilities. Similar to the work of \cite{deng2021labels} they trained a linear regression model for accuracy prediction. As input features to the regression model they used (separately) DoC, DoE, and distance metrics, \emph{e.g.},  Frech\'et Distance, MMD. They demonstrated that standard measures of distributional distance are often ineffective at predicting changes in accuracy when faced with natural shifts in data distribution. By treating DoC as a feature, they obtain regression models that significantly outperform  other methods. 

\subsection{Segmentation Model Performance Estimation}
To predict segmentation performance in the absence of ground truth labels \cite{valindria2017reverse} propose reverse classification accuracy (RCA). Specifically, a classifier is trained on a single image with its predicted segmentation (where the segmentation prediction of this single image is derived from the original segmentation model) serving as pseudo ground truth (GT). The method is built on the assumption that if the segmentation quality for a new image is high, then the RCA classifier trained on the predicted segmentation used as pseudo GT will perform well at least on some of the images in the reference database. Similarly, if the segmentation quality is poor, the classifier is likely to perform poorly on the reference images. The Reverse classifiers that are used in the study are Atlas forests \citep{zikic2014encoding}, the CNN model DeepMedic \citep{kamnitsas2017efficient} with a decreased amount of filters and consequently parameters to reduce overfitting to the single image input, and Atlas-based label propagation \citep{bai2013probabilistic}. 
It is important to note that the reference database used for evaluation can be the same as, or different from, the training database employed for training, cross-validation, and fine-tuning the original segmentation method. For measuring the segmentation accuracy the maximum Dice score coefficient value that is found across all reference images is used. They found that Atlas Forests and in particular, Single-Atlas label propagation yield accurate predictions (in terms of mean absolute error and correlation between predicted and true dice score coefficient) for different segmentation methods.




\section{Discussion}

In Sections~\ref{sec:methods_bias_encoding},  \ref{sec:methods_datadrift} and  \ref{sec:data_drift_harmfulness_assessment} we discuss methods to ensure reliable ML model deployment and maintain consistent prediction performance over time in clinical settings. Most of the methods presented have been tested on medical imaging data, while some, particularly those in Sect.~\ref{sec:methods_datadrift_imaging} (Drift Detection) and Sect.~\ref{sec:drift_severity_assessment},   \ref{sec:classifier_accuracy_estimation} (Performance Estimation), still need to be evaluated on medical imaging data. To ease comparison between the discussed approaches we have summarised their problem domain in Table~\ref{tab:feature_comparison}. 

The Bias Encoding Assessment methods outlined in Sect.~\ref{sec:bias_encoding_assessment} can help practitioners gain transparency regarding bias encoding in pre-trained medical imaging models and to derive tailored bias mitigation strategies. Beyond assessing bias through output prediction disparities across sensitive subgroups (sex, race, age), examining the inner workings of bias encoding offers insights into how sensitive attributes are inter-related with each other and related with the primary task (disease detection). As \cite{brown2023detecting} and \cite{glocker2023algorithmic} have demonstrated, encoding of sensitive attributes does not necessarily mean that those features are used for the primary task prediction. If bias is present, the impact of these sensitive attributes on model performance should be analysed (\emph{e.g.}, with the method proposed by \cite{brown2023detecting}) to distinguish between scenarios a and b as shown in Figure~\ref{fig:shortcuts_figure}.

Further research is required to determine whether the findings of \cite{glocker2023risk} on bias encoding in chest radiography foundation models also apply to other foundation models. Given the high popularity of foundation models in the medical field \citep{azad2023foundational} and the foundation models' use as a basis for other downstream task applications represents a key concern since any inherent biases of the foundation models might be  inherited by all models that are fine-tuned on them \citep{bommasani2021opportunities}. Therefore, it is crucial to tackle and alleviate biases in foundation models to guarantee fairness, inclusiveness, and ethical development within the medical domain \citep{azad2023foundational}.

Determining the root causes of performance disparities remains challenging and will require future research. The presence of multiple potential sources of bias, including selection bias, annotation bias, and algorithmic bias -- both demographic and non-demographic -- makes it challenging to pinpoint the exact factors contributing to the disparities. Additionally, the interconnections between these biases complicate causal bias determination.

In addition to efforts aimed at mitigating bias in models trained on medical imaging data \citep{dinsdale2021deep, correa2021two, yang2024limits}, it is crucial to collect a representative dataset to prevent the model from inheriting dataset bias. Attention should also be given to the labelling process, with procedures in place to resolve multi-annotator label disagreements, ensuring unbiased labels.

When developing models, there is an inherent trade-off between tailoring a model to the specific deployment population -- while risking bias from the under- or overrepresentation of certain subgroups -- and creating a model based on a more representative dataset for the broader population (using \emph{e.g.} reweighing strategies). The former approach may not generalise well to hospitals in other countries with different patient demographics, while the latter may lack the specificity needed for the deployment population, requiring fine-tuning for optimal performance.

Most of the drift detection methods discussed in Section~\ref{sec:methods_datadrift_imaging} were evaluated on simulated data drifts. Thus, future research is necessary to assess these methods on real data drifts. However, this poses a challenge, as, given we are mostly relying on temporal imaging data from a single source, it is inherently difficult to determine when a true statistically significant shift has occurred, which is essential for validating the effectiveness of these methods. 


For implementing drift detection methods, it is crucial to set the intervals at which drift detection will be applied, thereby determining the appropriate number of images needed for the target dataset. \cite{kore2024empirical} found that increasing the number of images in the target set improves sensitivity. However, this may not be feasible for smaller institutions or ML applications that handle relatively few cases, such as rare diseases. Thus, institutions may need to balance the frequency of drift detection with its sensitivity.

When a statistically significant data drift was detected and performance deterioration was estimated, expert labelling is necessary to annotate the data collected between the original model deployment and the drift detection timepoint. This step is crucial to confirm whether there was an actual decline in performance. Then retraining and re-validation of the model is necessary. 

An intriguing area of study arises when drift is detected without corresponding performance deterioration. In such cases, analysing which covariates have changed (\emph{e.g.}, patient demographics, scanner type) through exploratory analysis can provide insights into which populations the model remains generalizable to. However, this analysis must be approached with caution, as, even if generalizability to certain populations is observed, there is no guarantee that the model is generalizable to populations with similar distributions, since unknown confounding factors may still be present.

In this work, we primarily focus on methods for \emph{detecting} bias and data drifts, leaving the discussion of automated mitigation strategies for future exploration. Alternatives to re-labelling, re-training, and re-evaluation are not covered in depth. However, we propose two distinct pathways for addressing drift automatically: (a) domain adaptation techniques, \emph{e.g.}, \citep{kamnitsas2017unsupervised,ouyang2019data,ouyang2022causality,ouyang2022self,guan2021domain,chen2019synergistic}, which modify inputs to align them with a target distribution where a trained model remains effective, and (b) out-of-distribution detection~\cite{yang2024generalized,zimmerer2022mood,tan2021detecting,schluter2022natural,naval2024disyre,baugh2023many,p2023confidence,baugh2022nnood} coupled with biased model selection, where models tailored to specific populations are chosen based on explicit or implicit patient class parameters. The former has already shown promising results when transferring tasks across modalities~\citep{ouyang2022causality}, while the latter could be a straightforward option for production environments, though it necessitates rigorous validation for each intended-use sub-population.

\section{Conclusion}

In this survey, we reviewed current approaches for bias encoding assessment and data drift detection in disease detection models trained on medical imaging data. Additionally, we explored methods for estimating model accuracy following the detection of statistically significant data drift, particularly in situations where ground truth labels are unavailable due to the lack of timely expert annotations. These techniques may become  essential for assessing the impact of drift and ensuring reliable model performance.

Enhancing the reliability of ML models is critical for improving patient safety and supporting the seamless integration of these models into clinical practice. This need is further underscored by the introduction of new AI regulations, such as the EU AI Act~\citep{consilium} or California's AI bills AB 2013 and SB 1047~\citep{california_sb1047_2023,california_ab2013_2023}, which impose stringent transparency requirements on high-risk AI systems, including those used in medical devices. High-risk AI systems must be sufficiently transparent to allow users to interpret their outputs accurately and address potential risks, such as biases that could lead to discriminatory outcomes. Additionally, information regarding potential risks of discrimination should be made available to help mitigate health and safety concerns during the AI application process~\citep{eur_lex}.

By providing this comprehensive overview, we aimed at equipping practitioners with the tools and insights necessary to deploy unbiased ML models that maintain consistent predictive performance over time, fostering safer and more equitable use in clinical settings.

\section*{Acknowledgments}

The authors gratefully acknowledge the scientific support
provided by the ERC - project MIA-NORMAL 101083647,  DFG 513220538, 512819079, and by the state of Bavaria (HTA).
HPC resources where needed were provided by the Erlangen National High Performance Computing Center (NHR@FAU) of the Friedrich-Alexander-Universität Erlangen-Nürnberg (FAU) under the NHR projects b143dc and b180dc. NHR funding is provided by federal and Bavarian state authorities. NHR@FAU hardware is partially funded by the German Research Foundation (DFG) – 440719683. 
During the preparation of this work the authors used overleaf, dict.cc, and ChatGPT 4o in order to provide correct spelling and grammar and as thesaurus. After using this tool/service, the authors reviewed and edited the content as needed and take full responsibility for the content of the publication. No parts of these manuscript have been generated by a language model.

\bibliographystyle{model2-names.bst}\biboptions{authoryear}
\bibliography{Bibliography}

\newcommand{\cmark}{\textcolor{ForestGreen}{\ding{51}}}
\begin{table*}[ht]
\centering
\rowcolors{2}{gray!25}{white}
\vspace{0.3cm}
\begin{tabular}{@{}lccccccccccccccccc@{}}
 & & \multicolumn{3}{c}{Bias Assessment} & & \multicolumn{5}{c}{Data Drift Detection} & &
 \multicolumn{3}{c}{Harmfulness Evaluation} & & \multicolumn{2}{c}{Application} \\
 \cmidrule{3-5} \cmidrule{7-11} \cmidrule{13-15} \cmidrule{17-18}
\textbf{Approach}        &  & 
\rotatebox{90}{\parbox{3cm}{Demographic \\ Confounder}}
& 
\rotatebox{90}{\parbox{3cm}{Non-Demographic  Confounder}}
 &
 \rotatebox{90}{\parbox{3cm}{Synthetic Bias \\ Datasets}}
 & &  \rotatebox{90}{\parbox{3cm}{Drift Detection}} &  \rotatebox{90}{\parbox{3cm}{Prevalence Drift}} & 
  \rotatebox{90}{\parbox{3cm}{Shift Detection}} & 
  \rotatebox{90}{\parbox{3cm}{Synthetic Drift}} & 
  \rotatebox{90}{\parbox{3cm}{Natural Drift}} & &
 \rotatebox{90}{\parbox{3cm}{Drift Severity}} &
 \rotatebox{90}{\parbox{3cm}{Classifier \\ Accuracy}} &
 \rotatebox{90}{\parbox{3cm}{Segmentation \\ Accuracy}} & &
 \rotatebox{90}{\parbox{3cm}{RGB Image Data}} & 
 \rotatebox{90}{\parbox{3cm}{Medical Imaging \\ Data}} \\ 
\midrule
\cite{glocker2023algorithmic} &  &  
\cmark &   & \cmark & & 
  &   &   &   &   &   & 
&&&& \cmark & \cmark \\
\cite{glocker2023risk} &  &  \cmark &   &  & & &   &   &   &   &   & &&&& \cmark & \cmark \\
\cite{piccarra2023analysing}  &  &  \cmark &   &  & & &   &   &   &   &   & &&&& \cmark & \cmark \\
\cite{brown2023detecting} &  &  \cmark &   &  & & &   &   &   &   &   & &&&& \cmark & \cmark \\
\cite{boland2024there}  &  &  &  \cmark &  & & &   &   &   &   &   & &&&& \cmark & \cmark \\
\cite{wachinger2019quantifying}  &  & \cmark &  \cmark &  & & &   &   &   &   &   & &&&& \cmark & \cmark \\
\cite{stanley2023flexible}  &  &  &  & \cmark & & &   &   &   &   &   & &&&& \cmark & \cmark \\
\cite{stanley2024towards}  &  &  &  & \cmark & & &   &   &   &   &   & &&&& \cmark & \cmark \\
\cite{rabanser2019failing}  &  &  &  &  &  & \cmark &   & &  \cmark  &   &   & \cmark & &&& \cmark &  \\
\cite{ovadia2019can}  &  &  &  &  &  & \cmark &   & &  \cmark  &   &   & &&&& \cmark &  \\
\cite{kore2024empirical}  &  &  &  &  &  & \cmark &   & &  \cmark  &  \cmark &   & &&&& \cmark & \cmark \\
\cite{koch2024distribution}  &  &  &  &  &  & \cmark &   & &  \cmark  &   &   & &&&& \cmark & \cmark \\
\cite{merkow2023chexstray}  &  &  &  &  &  & \cmark &   & &  \cmark  &   &   & &&&& \cmark & \cmark \\
\cite{stacke2020measuring}   &  &  &  &  &  & \cmark &   & &  \cmark  &   &   & &&&& \cmark & \cmark \\
\cite{roschewitz2023automatic}   &  &  &  &  &  & \cmark & \cmark  & &  \cmark  &   &   & &&&& \cmark & \cmark \\
\cite{guo2023medshift}   &  &  &  &  &  &  &   & \cmark &  \cmark  &   &   & &&&& \cmark & \cmark \\
\cite{deng2021labels}     &  &  &  &  &  &  &   &  &    &   &   & & \cmark &&& \cmark &  \\
\cite{press2024entropy}   &  &  &  &  &  &  &   &  &    &   &   & & \cmark &&& \cmark &  \\
\cite{guillory2021predicting}    &  &  &  &  &  &  &   &  &    &   &   & & \cmark &&& \cmark &  \\
\cite{valindria2017reverse}  &  &  &  &  &  &  &   &  &    &   &   & &  & \cmark && \cmark &  \cmark\\
\bottomrule
\end{tabular}
\caption{Comparison of Bias Encoding Assessment, Data Drift Detection, Harmfulness Estimation, and Application in Medical Imaging Studies. }
\label{tab:feature_comparison}
\end{table*}


\end{document}